\def\year{2022}\relax
\documentclass[letterpaper]{article} 
\usepackage{aaai23}  
\usepackage{times}  
\usepackage{helvet}  
\usepackage{courier}  
\usepackage[hyphens]{url}  
\usepackage{graphicx} 
\usepackage{natbib}  
\usepackage{caption} 
\DeclareCaptionStyle{ruled}{labelfont=normalfont,labelsep=colon,strut=off} 
\usepackage{algorithm}
\usepackage{algorithmic}
\usepackage{amssymb}
\usepackage{amsmath}
\usepackage{booktabs}
\usepackage{newfloat}
\usepackage{listings}
\floatstyle{ruled}
\newfloat{listing}{tb}{lst}{}
\floatname{listing}{Listing}
\nocopyright
\title{See What You See: Self-supervised Cross-modal Retrieval of Visual Stimuli \\
from Brain Activity}
\author{
}
\affiliations{

    Zesheng Ye, \textsuperscript{\rm 1}\thanks{University of New South Wales}
    Lina Yao, \textsuperscript{\rm 1, }\textsuperscript{\rm 2}\thanks{CSIRO's Data 61}
    Yu Zhang, \textsuperscript{\rm 3}\thanks{Lehigh University}
    Sylvia Gustin \textsuperscript{\rm 1, }\textsuperscript{\rm 4}\thanks{Neuroscience Research Australia}

}

\usepackage{bibentry}

\begin{document}

\maketitle

\begin{abstract}
    Recent studies demonstrate the use of a two-stage supervised framework to generate images that depict human perception to visual stimuli from Electroencephalography~(EEG), referring to EEG-visual {\it reconstruction}.
    They are, however, unable to ``reproduce'' the exact visual stimulus, since it is the human-specified annotation of images, not their data, that determines what the synthesized images are.
    Moreover, synthesized images often suffer from noisy EEG encodings and unstable training of generative models, making them hard to recognize.
    Instead, we present a single-stage EEG-visual {\it retrieval} paradigm where the data of two modalities are correlated, as opposed to their annotations, allowing us to recover the exact visual stimulus for an EEG clip.
    Specifically, we maximize the mutual information between the EEG encoding and associated visual stimulus through optimization of a contrastive self-supervised objective, leading to two additional benefits.
    One, it enables EEG encodings to handle visual classes beyond seen ones during training, since learning is not directed at class annotations.
    In addition, the model is no longer required to generate every detail of the visual stimulus, but rather focuses on cross-modal alignment and retrieves images at the instance level, ensuring distinguishable model output.
    Empirical studies are conducted on the largest single-subject EEG dataset that measures brain activities evoked by image stimuli.
    We demonstrate the proposed approach completes an instance-level EEG-visual {\it retrieval} task, i.e., to report the exact visual stimulus which existing methods cannot.
    We also examine the implications of a range of EEG and visual encoder structures.
    Furthermore, for a mostly studied semantic-level EEG-visual {\it classification} task, despite not using class annotations, the proposed method outperforms state-of-the-art supervised EEG-visual {\it reconstruction} approaches, particularly on the capability of open class recognition.
\end{abstract}

\section{Introduction}
Vision, a core capability of human intelligence, is regarded as one of the most critical senses for obtaining information from the environment, and has long been the focus of a multidisciplinary research area aiming to understand how the brain responds to external visual stimuli.
The convergence on hardware and software of Brain-Computer Interface has advanced its utility in healthcare and smart living.
The benefits of such a mind reading, especially for those with disabilities in daily communication, will eventually accrue to the general public, if the human brain's responses to visual stimuli can be decoded through computational means.

\begin{figure}
    \centering
    \includegraphics[width=\linewidth]{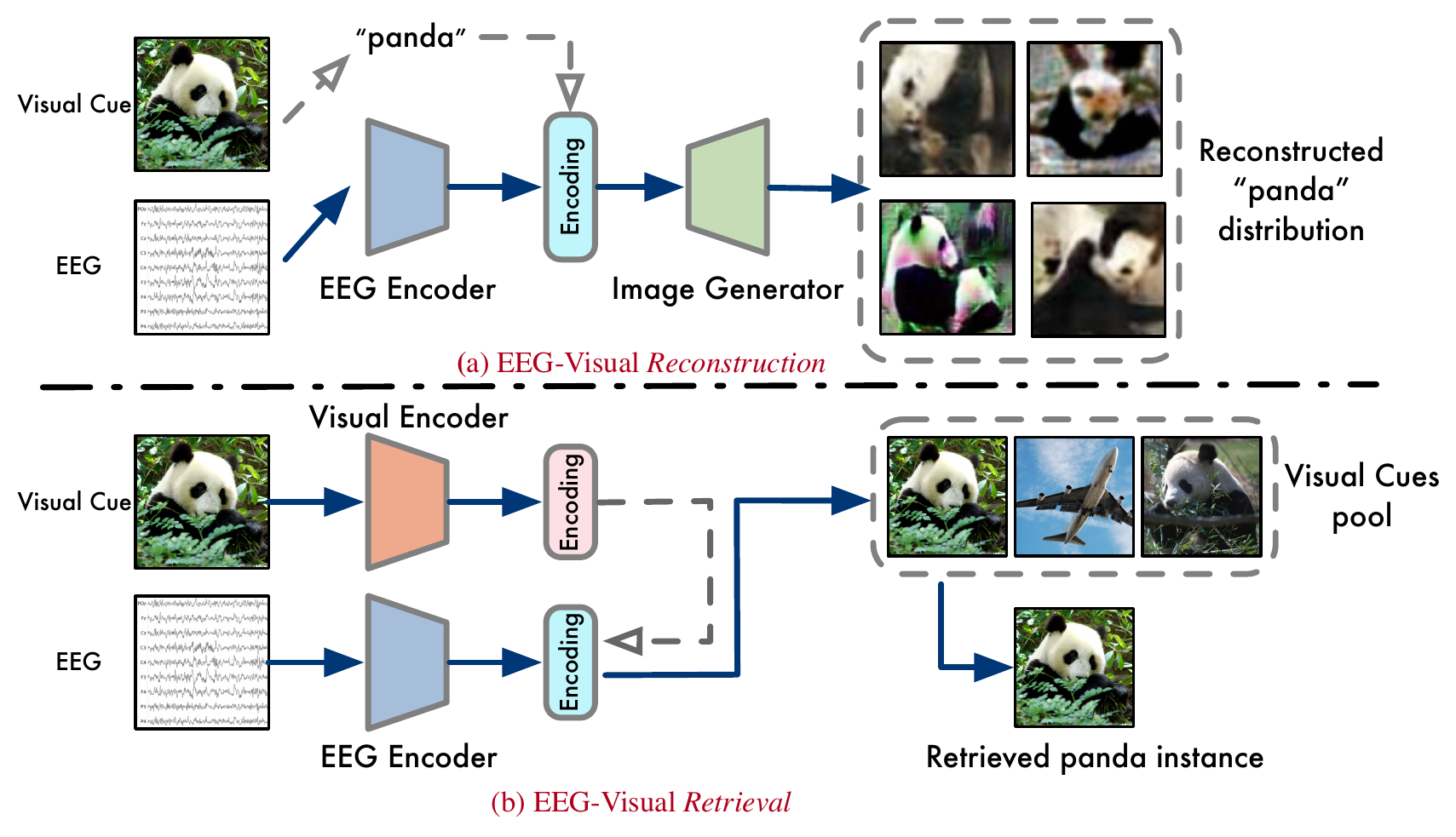}
    \caption{(a) visual {\it reconstruction} vs. (b) visual {\it retrieval}.}
    \label{fig: 1_concept}
\end{figure}

Machine learning in the last decade has significantly contributed to the progress on Electroencephalography~(EEG) recognition, thanks to the powerful expressiveness of deep neural networks~(DNNs).
A number of applications, such as intention prediction with EEG signals~\cite{zhang2019convolutional}, are cast into a supervised learning paradigm, by which the mapping from EEG data to the corresponding semantic class are encoded into a heuristic model, typically parameterized by DNNs.
Despite this paradigm works well in these {\it single-modal} EEG recognition tasks where the model only receives EEG as input while the output is manually defined semantic classes, it may struggle in tasks involving more than one modality, e.g., decoding visual stimuli from EEG signals.

It is challenging to recognize patterns in either EEG signals or visual stimuli, let alone both simultaneously, due to low signal-noise ratio of EEG data and high dimensionality of image data.
Studies have consistently followed the {\it single-modal} setting in existing EEG-to-vision literature.
That is, EEGs are encoded in accordance with visual classes, but not with actual images.
\cite{spampinato2017deep} initialize a task coined EEG-visual {\it recognition}, where they randomly select 50 images of 40 classes from ImageNet~\cite{russakovsky2015imagenet} and use them as visual cues shown to the subjects, whose brain activities were monitored by EEG during viewing of these visual cues.
Every individual EEG clip is mapped into a discriminative encoding that corresponds to one of 40 classes.
Built upon EEG-visual {\it recognition}, another line of EEG-to-vision studies aim at synthesizing the visual stimulus that evoke brain activities, known as EEG-visual {\it reconstruction}.
It has typically been implemented using a two-stage encoder-decoder framework, in which discriminative EEG encodings obtained from {\it recognition} task is extended to a generative decoder, so that the synthesis of images belonging to a particular visual class can be guided by the EEG encodings.
Fig.~\ref{fig: 1_concept}~(a) illustrates the pipeline of EEG-visual {\it reconstruction}.
\cite{kavasidis2017brain2image} deploy a two-layer Recurrent Neural Network~(RNN) to extract EEG encodings, and generate images with conditional Generative Adversarial Network~(GANs)~\cite{mirza2014conditional}, but using EEG encodings as class conditions.
It is primarily focused on improving either the encoder~(in classification) or decoder~(in generation) in the following works.
\cite{jiao2019decoding} concatenate EEG encodings with image features to provide a visually-guided EEG encoder.
\cite{tirupattur2018thoughtviz} apply a Mixture-of-Gaussian layer to reduce the data needed to train GANs.
In~\cite{fares2020brain}, both image features and EEG encodings are conditioned to GAN.
\cite{khare2022neurovision} progressively augment GAN with EEG features to improve the quality of synthesized images.

It is still overly optimistic to draw conclusions from these methods on visual {\it reconstruction}, because the generation of images is based on the semantic class, rather than the actual content of each individual visual stimulus.
If, for instance, the subject was looking at an image of a panda and the EEG was recorded, what these EEG-visual {\it reconstruction} models could do at best, is producing a bunch of images that of the visual class ``panda", but not necessarily the same panda viewed by this subject.
This problem arises from class-specific supervision not enabling the model to correlate the EEG modality with the visual modality.
It is thus difficult to obtain the expected ``instance-to-instance'' mapping from an EEG clip to the specific visual cue, because generative models essentially learn a ``distribution-to-distribution" mapping from noises to images~\cite{qiao2020biggan}.

Moreover, GANs have been found to struggle in normal image generation due to modes collapse and failure to converge~\cite{arjovsky2017wasserstein}.
Not to mention the EEG encoding used as class-specific guidance in EEG-to-vision tasks is far more noisy than ground-truth, making it hardly to generate sharp and distinguishable images.

To fullfil this gap, we propose to learn EEG-visual correspondence directly by maximizing the mutual information between the encoding of each EEG clip with the encoding of corresponding visual stimulus, in a self-supervised manner, namely {\bf EEG}-{\bf Vis}ion {\bf C}ross-{\bf M}odality {\bf R}etrieval~(EEGVis-CMR).
Particularly, we apply a contrastive objective based on the fact that every EEG clip is naturally paired with a visual cue, thereby maximizing intra-pair similarity and minimizing inter-pair similarity.
Ideally this method leads to finding a representation space, where an EEG encoding is located closest to image encoding of the corresponding visual cue, such that visual stimulus can be retrieved exactly for an EEG clip.
The merits that distinguish the proposed self-supervised {\it retrieval} paradigm from previous supervised {\it reconstruction}-based studies are three-fold.
First, it attains ``instance-to-instance'' EEG-image mapping while previous works do not, because of inherent limitation of current generative means;
Second, it ensures distinguishable model output by imposing instance-level cross-modal alignment.
This eliminates the need to train a decoder focusing on pixel-level generation but may not necessary to that task, i.e., reproducing the image that the subject viewed.
Third, it enables EEG encodings to extend beyond a limited set of visual classes, since cross-modal relationships are learned under the supervision of naturally formed EEG-image pairs, other than manual class annotations.
Empirical studies on~\cite{ahmed2021object}, the largest single-subject EEG dataset regarding visually-evoked tasks, shot that EEGVis-CMR performs competitively against state-of-the-arts.
In summary:
\begin{itemize}
    \item To our knowledge, we are the first to propose a contrastive self-supervised cross-modal learning approach that achieves ``instance-to-instance'' mapping from EEGs to visual stimulus, without using visual class annotations.
    \item Accordingly, we present a EEG-visual {\it retrieval} protocol to evaluate if the model is capable of recovering an accurate image instance given an EEG clip, in view of the limitations of previous {\it reconstruction}-based studies.
    \item We examine different EEG and visual encoders to investigate if more discriminative encodings can improve the model performance on EEG-visual {\it retrieval}.
    \item We compare EEGVis-CMR against the model outputs of state-of-the-arts on EEG-visual {\it classification}, showing that EEGVis-CMR outperforms the supervised {\it reconstruction} counterpart, even without class supervision. Moreover, EEGVis-CMR is shown to handle open-class visual stimuli with improved generalization ability empowered by self-supervision on the instance level.
\end{itemize}

\section{Problem Definition}
We briefly introduce EEG-visual brain activity evoking experiment process and formulate the EEG-visual {\it retrieval} problem.
Let $\mathcal{D}_{I} = \{ \mathbf{I}_{i, j} \}_{i=1:O, j=1:C}$ denote the dataset of $N$ images as visual cues, where there are $C$ different visual classes, and $O$ images for each.
During each session, the subject views one of $N = C \times O$ image stimuli, lasting for $L$-seconds, while an EEG helmet with $V$ sensory nodes is used to record the brain activity as an EEG clip.
Let $\mathbf{e}_{v \in [1, V]}$ be the recording sequence of a sensor.
Then each clip results in $\mathbf{e}_{v \in [1, V]} = [ e_{v}^{(1)}, e_{v}^{(2)}, \dots, e_{v}^{(T)} ] \in \mathbb{R}^{T}$ through $T = L \times f$ timesteps, where $f$ is the sampling frequency and $x_{v}^{(t)}$ is the measurement of the $v$-th node at $t$-th timestep.
The measurements of all sensors for an EEG clip is represented by $\mathbf{E} = \{ \mathbf{e}_{v} \}_{v \in [1, V]} \in \mathbb{R}^{V \times T}$.
Hence, there are $N$ clips recorded,
where each EEG clip $\mathbf{E}_{i, j}$ is paired with a visual cue $\mathbf{I}_{i, j}$, respectively.

EEG-visual {\it retrieval} aims to find the correct image paired with a query EEG clip.
Following common retrieval settings, the model matches the query EEG with every candidate image, and reports the image with the highest similarity.

\begin{figure*}
    \centering
    \includegraphics[width=\textwidth]{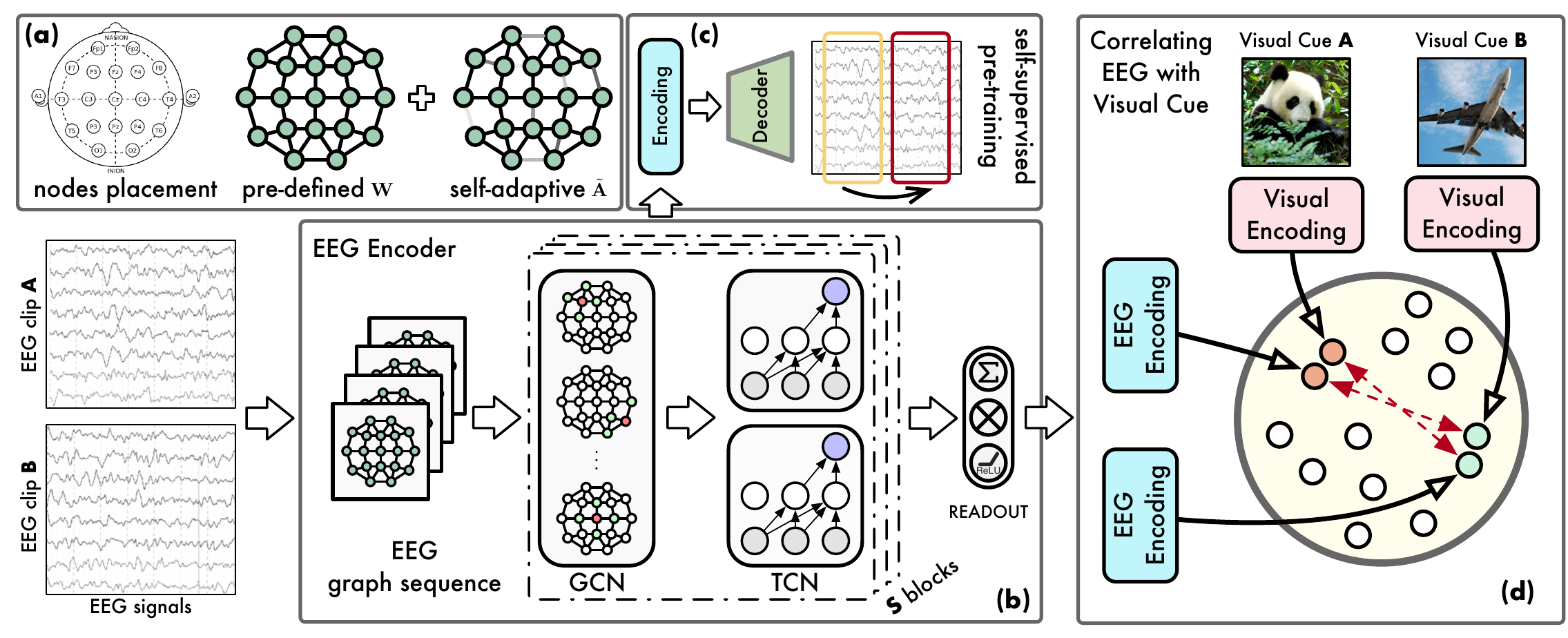}
    \caption{Method Illustration. {\bf (a)} The EEG sensory nodes placement based on 10-20 internationl system, the pre-defined adjacency $\mathbf{W}$ and self-adaptive adjacency $\tilde{\mathbf{A}}$. {\bf (b)} Temporal Graph Convolution Network as EEG encoder, taking each of EEG clips as input and producing corresponding EEG encoding. {\bf (c)} Self-supervised pre-training for EEG encoding, taking historical $T$ signal measurements as input and making prediction on the next $T^{\prime}$ timesteps. {\bf (d)} Contrastive learning for correlating EEG encoding with the visual encoding by maximizing the mutual information of correct EEG-image pairs.}
\end{figure*}

\section{Approach}
\subsection{Overview}
With $N$ paired EEG-image tuples as input, EEGVis-CMR first extracts the features of both modalities using domain-specific encoders.
Following the extraction of EEG and visual encodings, they are transformed into joint multi-modal encodings using two linear projection matrices, respectively.
EEGVis-CMR aims to create a shared latent representation space, in which each EEG encoding is drawn closer to the paired visual encoding while being pushed away from other $N-1$ EEG encodings and $N-1$ visual encodings.
This is optimized with respect to an instance-level discriminative objective.
In such a space, EEG encodings can be compared to every visual encoding to derive the model output.

\subsection{Encoding Visual Cues}
Denoted by $f_{\text{img}}$ , any DNN-based parametric model can be used to encode input raw images into visual features $\mathbf{z}_{\mathbf{I}} = f_{\text{img}}(\mathbf{I})$,
where $\mathbf{I}$ is the input image.
In EEGVis-CMR, the visual encodings are factual supervision used to encode EEGs.
This leads us to wonder if more powerful visual encoders can improve performance in this cross-modal task as expected.
We attempt with several DNN-based visual encoders, including AlexNet~\cite{krizhevsky2012imagenet}, ResNet~\cite{he2016deep}, and ViT~\cite{dosovitskiy2020image}, all have been pre-trained with ImageNet.

\subsection{Encoding EEGs}
Previous EEG recognition studies generally use recurrent neural networks (RNNs) and convolutional neural networks~(CNNs) to encode the spatio-temporal dependencies within EEG signals~\cite{zhang2018cascade,zhang2020motor}.
However, CNNs are unable to measure non-euclidean geometry and dynamic connectivity in brain networks under the assumption of Euclidean structure~\cite{tang2021self}.
RNNs, on the other hand, may struggle to model long-range temporal dependencies due to vanishing gradient.
As such, we apply a temporal graph convolution network~(TGCN) to encode EEG signals with a) self-adaptive graph convolution~(GCN) to capture dynamic brain connectivity and spatial relationship among EEG sensory measurements;
and b) gated temporal convolution~(TCN) to handle temporal dependencies of long sequences.
Specifically, the EEG encoder~(TGCN) comprises an output layer and $S$ stacked TGCN blocks, where each block consists of a GCN and a TCN layer for modeling spatial and temporal correlations, respectively.
Except for the last one, all blocks have their outputs used as input to the next block.
A permutation-invariant READOUT function collects node features after the final TGCN block, and produces the spatio-temporal EEG features $\mathbf{z}_{\mathbf{E}} = f_{\text{EEG}}(\mathbf{E}$), defined by
\begin{equation}
    f_{\text{EEG}} := \texttt{READOUT} \left( \left\{ \mathbf{h}^{(s)} \mid
    s = 1 \dots S \, ; \mathbf{E} \right\} \right)
\end{equation}
where $\mathbf{E}$ denotes the EEG input, $\mathbf{h}^{(s)}$ is the output graph of $s$-th TGCN block.
Here \texttt{READOUT}$(\cdot)$ is the summation on node features, plus non-linear transformations.

\subsubsection{Representing Dynamic Connectivity of EEG Sensors.}
To model spatial relationship, we represent the EEG signal as a undirectional weighted graph $G = \left< \mathcal{V}, \mathcal{E}, \mathbf{W} \right>$, where $\mathcal{V}, \mathcal{E}$ are the node set (i.e., EEG nodes) and edge set, respectively.
There exists an edge between any pair of nodes $v_i, v_j \in \mathcal{V}$ if the Euclidean distance $\text{dist}(v_i, v_j)$ between is among the $k$-th lowest distance connecting to node $v_i$, plus self-loop.
$\mathbf{W} \in \mathbb{R}^{V \times V}$ denotes the adjacency matrix whose entry $\mathbf{W}_{ij} = 1$ if $v_j \in \mathcal{N}(v_i)$, otherwise 0, where $\mathcal{N}(v_i)$ represents $v_i$'s neighbors connected by edges.
Thus, the connectivity between nodes are encoded by the adjacency matrix $\mathbf{W}$, with entries of higher value indicate stronger connectivity, and vice versa.

To this end, we compute self-adaptive adjacency~\cite{wu2019wave} to dynamically measure brain connectivity by deriving spatial correlations among two randomly initialized trainable node embeddings $\mathbf{\Theta}_{\mathbf{X}}, \mathbf{\Theta}_{\mathbf{X}^{\prime}} \in \mathbb{R}^{V \times C}$, leading to an self-adaptive adjacency matrix,
\begin{equation}
    \tilde{\mathbf{A}} = \texttt{SoftMax}\left( \texttt{ReLU} \left( \mathbf{\Theta}_{\mathbf{X}} \, \mathbf{\Theta}_{\mathbf{X}^{\prime}}^{\top} \right) \right) \in \mathbb{R}^{V \times V}
\end{equation}
where {\tt ReLU} introduces sparsity into the matrix, and {\tt SoftMax} normalizes the values.

We apply graph convolution~\cite{welling2016semi} to extract the node features $\tilde{\mathbf{X}} \in \mathbb{R}^{V \times M}$, by aggregating and transforming the neighbor information for each node $v \in \mathcal{V}$, from a spatial-based interpretation.
As~\cite{tang2021self} show that spatial correlations among EEG channels can be modeled using DCRNN~\cite{li2018dcrnn_traffic}, which approximates graph convolution into a $K$-step diffusion process, we adapt a generalized form of DCRNN to obtain the node features,
\begin{equation}
    \tilde{\mathbf{X}} = \sum_{k=0}^{K-1} \left( \mathbf{P}^{k} \mathbf{X} \mathbf{\Theta}_{\mathbf{W}}^{k} +  \tilde{\mathbf{A}}^{k} \mathbf{X} \mathbf{\Theta}_{\mathbf{A}}^{k} \right)
\end{equation}
where $\mathbf{X} \in \mathbb{R}^{V \times D}$ is the input embedding of EEG signal on each time-slice $\mathbf{E}[:, t]$.
$\mathbf{P}^{k} = \mathbf{W} / \texttt{rowsum}(\mathbf{W})$ is the power series of random walk transition matrix at the $k$-th step, generalizing the pre-defined adjacency $\mathbf{W}$ under $K$-step diffusion.
$\mathbf{\Theta}_{\mathbf{W}}^{k}, \mathbf{\Theta}_{\mathbf{A}}^{k}$ are trainable parameters for modeling pre-defined spatial correlations and self-adaptive embedding dependencies, respectively.

\subsubsection{Capturing Long-range Temporal Dependencies.}
RNN-based approaches have effectively expanded their horizons thanks to the gate mechanism~\cite{cho2014gru} that controls information propagation through time.
Moreover, it is applicable to TCN~\cite{dauphin2017language,wu2019wave}.
To detail the formulation of TGCN block, we slightly omit the TGCN block index $\mathbf{h}^{(s)}$ and time index $\mathbf{X}^{(t)}$ for brevity.

Consider the embedding of an EEG clip $\mathbf{X} \in \mathbb{R}^{V \times D \times T}$ of $T$ timesteps, we adopt the gated TCN to model temporal dependencies from the spatial features $\tilde{\mathbf{X}} \in \mathbb{R}^{V \times M \times T}$ derived by self-adaptive GCN applied on each time-slice $\mathbf{X}[:, :, t]$,
\begin{equation}
    \mathbf{h} = \tanh \left( \mathbf{\Theta}_{\mathbf{b}} \star \tilde{\mathbf{X}} + \mathbf{b} \right) \odot \sigma \left( \mathbf{\Theta}_{\mathbf{c}} \star \tilde{\mathbf{X}} + \mathbf{c} \right)
\end{equation}
where $\mathbf{\Theta}_{\mathbf{b}}, \mathbf{\Theta}_{\mathbf{c}}, \mathbf{b}, \mathbf{c}$ are trainable TCN parameters, $\star$ is the convolution operator, $\sigma(\cdot)$ denotes Sigmoid function controlling the ratio of information passed to the next block, and $\odot$ refers to Hadamard production.

\subsubsection{Pre-training EEG Encoder.}
Self-supervised pre-training is shown to be effective at further improving model performance, by predicting next $T^{\prime}$ steps' measurements~\cite{tang2021self} given an input EEG clip with $T$ timesteps.

We thus perform task-irrelevant pre-training to obtain more robust EEG encodings by using an encoder-decoder architecture.
Given an EEG clip $\mathbf{E}^{1:T} \in \mathbb{R}^{V \times T}$, the encoder produces $\mathbf{z}_{\text{EEG}}$ which is then fed into to the decoder, leading to the estimation of next $T^{\prime}$ timesteps $\hat{\mathbf{E}}^{(T+1):(T+T^{\prime})} \in \mathbb{R}^{V \times T^{\prime}}$.
This brings in the self-supervised objective function with Mean Absolute Error,
\begin{equation}
    \mathcal{L}_{\text{pre}} =  (1 / T^{\prime}) \; \cdot \sum_{t=T+1}^{T+T^{\prime}} | \mathbf{E}^{(t)} - \hat{\mathbf{E}}^{(t)} |
\end{equation}
Noted that pre-training does not participate in the cross-modal learning pipeline of EEGVis-CMR and is optional.
We use the pre-trained parameters of encoder to produce EEG encodings $\mathbf{z}_{\mathbf{E}}$ in the downstream task.

\subsection{Correlating EEG with Visual Cues}
Ideally, a cross-modal learning model maps the features of both modalities into a shared latent metric space, so that the encodings of different data modalities are comparable.
While generative predictive methods are common practices to match the distributions of two modalities~\cite{gu2018look}, recent evidences indicate that contrastive representation learning could be more robust~\cite{tian2020rethinking} and more training-friendly~\cite{chen2020generative} than its predictive counterpart.
More importantly, contrastive learning imposes instance-level alignment rather than matching distributions as in generative approaches.
This allows exact retrieval of one instance, given a query sample from another modality.
Thus, inspired by~\cite{radford2021learning} which learns image features from natural language supervision, we propose to guide EEG encodings with image content, using a contrastive self-supervised objective.

Concretely, we are interested in finding the most similar visual encoding to a query EEG encoding.
Given $N$ pairs of EEG-image tuples $\left\{(\mathbf{E}_{i}, \mathbf{I}_{i}) \right\}_{i=1:N}$, we first obtain their domain-specific encodings, resulting in $N$ pairs of encoding tuples $\left\{(\mathbf{z}_{\mathbf{E}}^{i}, \mathbf{z}_{\mathbf{i}}^{i}) \right\}_{i=1:N}$.
These domain-specific encodings are then projected into a joint multi-modal space with transformation matrices $\mathbf{W}_{\mathbf{E}}$ and $\mathbf{W}_{I}$ plus non-linearity, for the respective modalities.
\begin{equation}
    \tilde{\mathbf{z}}_{\mathbf{E}} = \texttt{ReLU} (\mathbf{W}_{\mathbf{E}} \mathbf{z}_{\mathbf{E}}), \quad \tilde{\mathbf{z}}_{\mathbf{I}} =  \texttt{ReLU}(\mathbf{W}_{\mathbf{I}} \mathbf{z}_{\mathbf{I}})
\end{equation}
where $\tilde{\mathbf{z}}_{\mathbf{E}}$ and $\tilde{\mathbf{z}}_{\mathbf{I}}$ are EEG encodings and visual encodings in that space.
Following that, we create a $N \times N$ similarity matrix to denote the cross-modal pairwise similarities, where each row (column) indices an EEG clip (image), and the $(i, j)$-th entry refers to the similarity between the $i$-th EEG clip and the $j$-th image.
Recall that each EEG clip records the brain's response to only one visual stimulus in the experiment.
Thus the similarity matrix identifies all correctly paired EEG-image tuples by diagonal entries, i.e., $\text{sim}(\tilde{\mathbf{z}}_{\mathbf{E}}^{i}, \tilde{\mathbf{z}}_{\mathbf{E}}^{i})$, whereas the rest represent incorrect EEG-image pairs, e.g., $\text{sim}(\tilde{\mathbf{z}}_{\mathbf{E}}^{i}, \tilde{\mathbf{z}}_{\mathbf{I}}^{j})$ with $i \neq j$.
It is therefore expected to maximize the diagonal entries and minimize the others simultaneously.
This is equivalent to maximizing the mutual information between EEG encoding and corresponding visual encoding for each correct EEG-image pair, by optimizing for the InfoNCE loss~\cite{oord2018representation}.
\begin{equation}
    \begin{aligned}
        \mathcal{L} = \sum_{i=1}^{N} \log
         & \frac{  s(\tilde{\mathbf{z}}_{\mathbf{E}}^{i}, \tilde{\mathbf{z}}_{\mathbf{I}}^{i})  }{s(\tilde{\mathbf{z}}_{\mathbf{E}}^{i}, \tilde{\mathbf{z}}_{\mathbf{I}}^{i}) + \sum_{j=1, j \neq i}^{N}  s(\tilde{\mathbf{z}}_{\mathbf{E}}^{i}, \tilde{\mathbf{z}}_{\mathbf{I}}^{j}) +  s(\tilde{\mathbf{z}}_{\mathbf{E}}^{i}, \tilde{\mathbf{z}}_{\mathbf{E}}^{j})   } \\
         & \qquad \qquad \qquad \qquad \; \; + s(\tilde{\mathbf{z}}_{\mathbf{E}}^{j}, \tilde{\mathbf{z}}_{\mathbf{I}}^{I}) +  s(\tilde{\mathbf{z}}_{\mathbf{I}}^{i}, \tilde{\mathbf{z}}_{\mathbf{I}}^{j})
    \end{aligned}
\end{equation}
with
\begin{equation*}
    s(\mathbf{a}, \mathbf{b}) = \exp \left( \frac{ \mathbf{a}^{\top} \, \mathbf{b}  }{ || \mathbf{a} || \, || \mathbf{b} || \cdot \tau }   \right)
\end{equation*}
defines an exponential term of the cosine similarity of two input vectors,
where $\tau$ is the temperature coefficient determining how the hard negatives are focused on.
It is only the paired $i$-th visual encoding considered to be the positive sample for the $i$-th EEG encoding, whereas the remaining visual encodings and EEG encodings are negative samples of the query EEG.
In addition, we enforce that the similarity between different instances of the same modality should be minimized, to improve the efficiency of negative sampling.
By doing so, each EEG encoding is attracted by the postive visual encoding while being repelled from other negative $N - 1$ images and $N - 1$ EEGs, leading to a joint representation space where multi-modal encodings are uniformly distributed.

\section{Empirical Studies}
\subsection{Experimental Setup}
\subsubsection{Dataset.}
We verify the proposed approach using the largest publicly-accessible single-subject EEG dataset that measures brain activity evoked by visual stimuli~\cite{ahmed2021object}.
The subject is shown 40,000 images and recorded 40,000 EEG segments in response to those visual cues.
The 40,000 cues downloaded from ImageNet belong to 40 visual classes, each containing 1,000 images.
Each of the 40,000 EEG clips is recorded with an EEG helmet of 96 channels at 4,096 Hz, lasting for two seconds.
All the channels and full-length data are used for experiments.

\subsubsection{Dataset Splits.}
A series of EEG-vision studies~\cite{spampinato2017deep, kavasidis2017brain2image, tirupattur2018thoughtviz, jiao2019decoding, fares2020brain} have been criticized for causing label leakage from the training to the test set, because of incorrect experimental settings~\cite{li2020perils}.
To address this concern, 40,000 images were partitioned into 100 disjoint sets, each containing ten images representing each of 40 visual classes, ensuring that no EEG clips intersected between the training and test set.

For each run of 5-fold cross-validation, two sets are randomly selected as validation and test.
The remaining 98 sets are used to train the model.

\subsubsection{Data Preprocessing.}
Each EEG clip used in this study is pre-processed as follows.
We first down-sample each of the EEG clips to 1,024 Hz from 4,024 Hz.
Then, we apply Butterworth bandpass filter to keep signals between 55-95~Hz, which appear to be more closely associated with visual-evoked brain responses~\cite{palazzo2020decoding}.
We lastly normalize the EEG clip using the mean and standard deviation of the training data.

As the adopted visual encoders expect each image input to have a shape of 224 $\times$ 224, we scale the images accordingly, without further data augmentation operations.

\subsubsection{Metrics.}
The adopted performance metrics on EEG-visual {\it retrieval} include mean Reciprocal Rank~(MRR) and mean Average Precision~(mAP) as a ranking task.
The MRR measures how well the retrieval model ranks the unique correctly paired visual cue, whereas the mAP measures all the relevant visual cues, which refer to images belonging to the same semantic class as the correct matching.
As for EEG-visual {\it classification}, we use accuracy for evaluation since both EEG and image data are balanced.
\subsection{Instance-level EEG-Visual Retrieval}
\label{subsec: exp_fgvr}

Given an EEG clip, EEG-visual {\it retrieval} aims to find the correctly paired visual cue.
In this case, it is necessary to output exact rankings of visual instances than addressing semantic visual classes only.
We summarize the performances of EEGVis-CMR with a range of variants upon performing EEG-visual {\it retrieval}.
In addition, we investigate how negative sampling affects contrastive learning of EEGVis-CMR.

\begin{figure*}
    \centering
    \includegraphics[width=\textwidth]{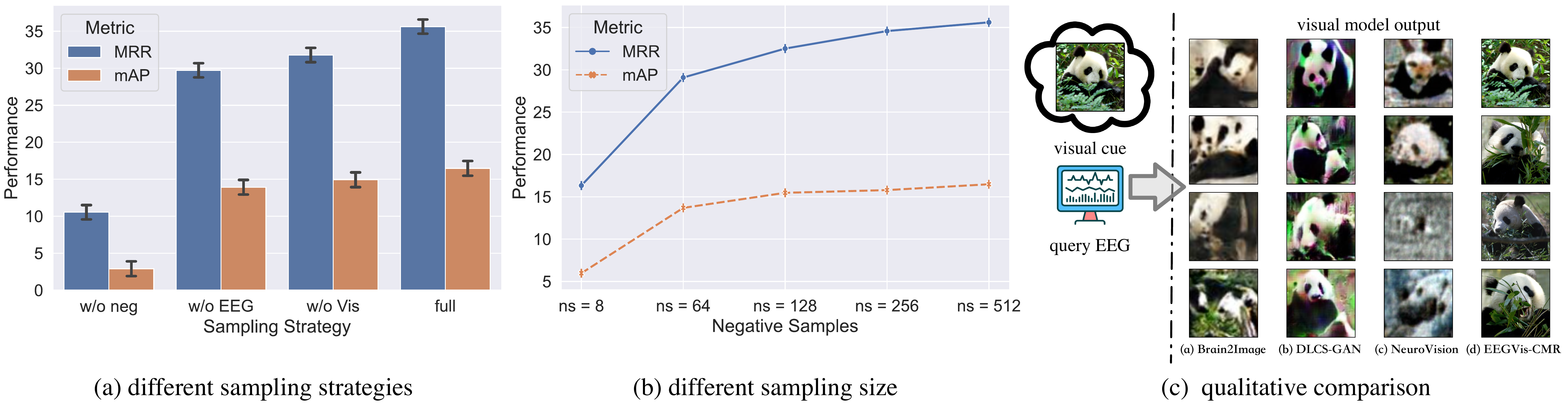}
    \caption{(a) Performance comparison of differenet negative sampling strategies. (b) Performance comparison of different negative sampling sizes. (c) Qualitative comparison of (a) Brain2Image, (b) DCLS-GAN, (c) NeuroVision, (d) EEGVis-CMR.}
    \label{fig: qualitative_ablation}
\end{figure*}

\subsubsection{Quantitative Results.}
\begin{table}[]
    \caption{Quantitative results of the proposed EEGVis-CMR on EEG-visual {\it retrieval}, measured by MRR and mAP. Variants of different visual encoders, EEG encoders are included. For visual encoder variants, AN, RN indicate AlexNet and ResNet-50, respectively. We also report the performance difference obtained with pre-training the EEG encoder. Mean and std are from 5-fold cross validation with the same random seed. The best results are {\bf bolded} and the second best are \underline{underlined}.}
    \begin{tabular}{l|l|l}
        \toprule
        {\bf method $\backslash$ metric} & {\bf MRR ($\times 10^{2}$)}   & {\bf mAP ($\times 10^{2}$)}   \\
        \toprule
        \multicolumn{3}{l}{{\it Visual Encoder variants}}                                                \\
        \midrule
        EEGVis-CMR (AN)                  & 25.82 $\pm$ 0.824             & 11.93 $\pm$ 0.261             \\
        EEGVis-CMR (RN)                  & 31.26 $\pm$ 1.252             & 14.54 $\pm$ 0.394             \\
        \midrule
        \multicolumn{3}{l}{{\it EEG Encoder variants}}                                                   \\
        \midrule
        EEGVis-CMR (RNN)                 & 27.55 $\pm$ 0.771             & 12.52 $\pm$ 0.229             \\
        EEGVis-CMR (CNN)                 & 28.86 $\pm$ 0.828             & 13.15 $\pm$ 0.333             \\
        \midrule
        EEGVis-CMR w/o pre               & \underline{32.79 $\pm$ 1.649} & \underline{15.68 $\pm$ 0.579} \\
        EEGVis-CMR                       & {\bf 35.59 $\pm$ 0.869}       & {\bf 16.49 $\pm$ 0.370}       \\
        \bottomrule
    \end{tabular}
    \label{table: retrieval}
\end{table}
We report the quantitative results of EEGVis-CMR on EEG-Visual {\it Retrieval} with Table~\ref{table: retrieval}.
Generally, EEGVis-CMR can yield reasonable retrieval results.
Both metrics show the same trend across the variants.
The model achieves higher MRR would also produce higher mAP, and vice versa.
It therefore can be concluded that the proposed contrastive self-supervised framework is capable of this EEG-vision cross-modality learning task.
Pre-training of EEG encoders is also shown to be effective empirically, as shown in the last two rows in Table~\ref{table: retrieval}.
We can observe performance differences on both metrics, with and without pre-training step.
Further interest is found in the fact that EEGVis-CMR is a unified framework, which can be configured with a variety of domain-specific encoders.

\subsubsection{Impact of Visual Encoder.}
In EEGVis-CMR, we use visual encodings to provide supervision to the EEG encodings.
Intuitively this is more informative than learning from limited semantic visual classes, but also results in noises than ground-truth labels.
The pre-trained AlexNet~\cite{krizhevsky2012imagenet}, ResNet-50~\cite{he2016deep}, and ViT~\cite{dosovitskiy2020image} are tested, each assuming a certain level of discriminative visual encoding.
Denote the variants with AlexNet, ResNet-50 and ViT as EEGVis-CMR~(AN), EEGVis-CMR~(RN) and EEGVis-CMR, respectively, we can attribute the improved performance to more powerful visual encoders, taking into account they share the same EEG encoder model and training policy.


\subsubsection{Impact of EEG Encoder.}
We also examine the performance of EEGVis-CMR with different EEG encoder structures.
As aforementioned, there are limitations to RNN-based and CNN-based encoders.
Our aim is to determine whether the devised GNN-based approach that incorporate non-euclidean structures, dynamic connectivity and temporal dependency is beneficial to the modeling.
We additionally take two baselines of EEG encoders into comparison, adopting a 2D temporal CNN and a two-layer bi-directional GRU with the design in~\cite{kavasidis2017brain2image}, denoted as EEGVis-CMR~(CNN) and EEGVis-CMR~(RNN).
The present GNN-based approach proves the most effective out of three with improvements on both metrics.

\subsubsection{Impact of Negative Sampling.}
We present comparisons of different negative sampling strategies and sizes for contrastive learning within EEGVis-CMR.
Fig.~\ref{fig: qualitative_ablation}~(a) compares four different strategies.
Observe that it yields the lowest results without negative samples.
Moreover, while using only either negative EEGs or visual encodings is feasible, the best performance is obtained when both negative embeddings are retained.
For Fig.~\ref{fig: qualitative_ablation}~(b), we find an upward trend in the model performance as the number of negative samples increases.
In this task, the larger of negative samples confirms the general hypothesis of contrastive learning.

\subsubsection{Qualitative Assessment.}
Furthermore, we showcase the qualitative results of EEGVis-CMR as well as three state-of-the-art {\it reconstruction}-based EEG-vision approaches, Brain2Image~\cite{kavasidis2017brain2image}, DCLS-GAN~\cite{jiao2019decoding}, NeuroVision~\cite{khare2022neurovision}.
The task-oriented perspective suggests instance-level alignment is more advantageous than distribution-level generation.
Fig.~\ref{fig: qualitative_ablation} shows a comparison of model outputs using a panda's image as visual cue.
The \textit{reconstruction}-based methods synthesize increasingly distinguishable images belonging to the visual class ``panda'', they fail to produce images identical to the visual stimuli.
EEGVis-CMR, on the other hand, can directly retrieve the exact paired image from those visual stimuli.


\subsection{Semantic-level EEG-Visual Classification}
\label{subsec: exp_cgvc}

\begin{table*}[htbp!]
    \caption{The EEG-visual {\it classification} results on all of 40 visual classes. The first 39 visual classes are evaluated under closed-set setting, whereas the last class~({\tt n13054560}) is left for open-set recognition. Results are from 5 different evaluation splits.}
    \label{tab: classification}
    \resizebox{\textwidth}{!}{%
        \begin{tabular}{l|llll||l|llll}
            \toprule
            {\bf Visual Class} & {\bf Brain2Image}  & {\bf DCLS-GAN}     & {\bf NeuroVision}  & {\bf EEGVis-CMR}               & {\bf Visual Class} & {\bf Brain2Image}        & {\bf DCLS-GAN}           & {\bf NeuroVision}        & {\bf EEGVis-CMR}               \\
            \midrule
            \midrule
            n02106662          & 0.029 $\pm$ 0.0178 & 0.163 $\pm$ 0.0249 & 0.058 $\pm$ 0.0336 & \underline{0.395 $\pm$ 0.1238} & n02951358          & 0.122 $\pm$ 0.0249       & 0.195 $\pm$ 0.0274       & 0.131 $\pm$ 0.0488       & \underline{0.509 $\pm$ 0.1294} \\
            n02124075          & 0.091 $\pm$ 0.0116 & 0.219 $\pm$ 0.0116 & 0.162 $\pm$ 0.1038 & \underline{0.507 $\pm$ 0.2107} & n02992529          & 0.039 $\pm$ 0.0155       & 0.271 $\pm$ 0.0152       & 0.041 $\pm$ 0.0416       & \underline{0.388 $\pm$ 0.2551} \\
            n02281787          & 0.141 $\pm$ 0.0117 & 0.311 $\pm$ 0.0117 & 0.184 $\pm$ 0.0737 & \underline{0.512 $\pm$ 0.1315} & n03063599          & 0.027 $\pm$ 0.0072       & 0.153 $\pm$ 0.0285       & 0.099 $\pm$ 0.1089       & \underline{0.369 $\pm$ 0.2213} \\
            n02389026          & 0.103 $\pm$ 0.0199 & 0.195 $\pm$ 0.0656 & 0.198 $\pm$ 0.0545 & \underline{0.578 $\pm$ 0.1116} & n03100240          & 0.186 $\pm$ 0.0269       & 0.297 $\pm$ 0.0819       & 0.238 $\pm$ 0.0603       & \underline{0.498 $\pm$ 0.1536} \\
            n02492035          & 0.272 $\pm$ 0.0379 & 0.297 $\pm$ 0.0808 & 0.308 $\pm$ 0.0487 & \underline{0.431 $\pm$ 0.1538} & n03180011          & 0.112 $\pm$ 0.0344       & 0.206 $\pm$ 0.0645       & 0.136 $\pm$ 0.0587       & \underline{0.463 $\pm$ 0.1439} \\
            n02504458          & 0.195 $\pm$ 0.0616 & 0.219 $\pm$ 0.0976 & 0.265 $\pm$ 0.0689 & \underline{0.414 $\pm$ 0.2307} & n04555897          & 0.131 $\pm$ 0.0495       & 0.186 $\pm$ 0.0682       & 0.227 $\pm$ 0.1092       & \underline{0.502 $\pm$ 0.1111} \\
            n02510455          & 0.181 $\pm$ 0.0674 & 0.258 $\pm$ 0.1185 & 0.274 $\pm$ 0.0331 & \underline{0.565 $\pm$ 0.1372} & n03272010          & 0.065 $\pm$ 0.0232       & 0.261 $\pm$ 0.0319       & 0.076 $\pm$ 0.0787       & \underline{0.485 $\pm$ 0.2276} \\
            n02607072          & 0.166 $\pm$ 0.0287 & 0.384 $\pm$ 0.0721 & 0.209 $\pm$ 0.0585 & \underline{0.623 $\pm$ 0.1794} & n03272562          & 0.074 $\pm$ 0.0157       & 0.196 $\pm$ 0.0732       & 0.172 $\pm$ 0.0648       & \underline{0.494 $\pm$ 0.1152} \\
            n02690373          & 0.227 $\pm$ 0.0518 & 0.433 $\pm$ 0.0522 & 0.327 $\pm$ 0.1264 & \underline{0.606 $\pm$ 0.0861} & n03297495          & 0.154 $\pm$ 0.0251       & 0.265 $\pm$ 0.0191       & 0.189 $\pm$ 0.0541       & \underline{0.561 $\pm$ 0.1769} \\
            n02906734          & 0.071 $\pm$ 0.0373 & 0.273 $\pm$ 0.0292 & 0.101 $\pm$ 0.0676 & \underline{0.518 $\pm$ 0.2024} & n03376595          & 0.061 $\pm$ 0.0194       & 0.167 $\pm$ 0.0266       & 0.115 $\pm$ 0.0394       & \underline{0.575 $\pm$ 0.1187} \\
            \midrule
            \midrule
            {\bf Visual Class} & {\bf Brain2Image}  & {\bf DCLS-GAN}     & {\bf NeuroVision}  & {\bf EEGVis-CMR}               & {\bf Visual Class} & {\bf Brain2Image}        & {\bf DCLS-GAN}           & {\bf NeuroVision}        & {\bf EEGVis-CMR}               \\
            \midrule
            \midrule
            n03445777          & 0.079 $\pm$ 0.0238 & 0.213 $\pm$ 0.0313 & 0.161 $\pm$ 0.0774 & \underline{0.408 $\pm$ 0.1091} & n03888257          & 0.131 $\pm$ 0.0214       & 0.333 $\pm$ 0.0237       & 0.241 $\pm$ 0.0669       & \underline{0.546 $\pm$ 0.0941} \\
            n03452741          & 0.114 $\pm$ 0.0323 & 0.261 $\pm$ 0.0422 & 0.185 $\pm$ 0.0559 & \underline{0.461 $\pm$ 0.1649} & n03982430          & 0.186 $\pm$ 0.0121       & 0.285 $\pm$ 0.0198       & 0.234 $\pm$ 0.1058       & \underline{0.481 $\pm$ 0.0689} \\
            n03584829          & 0.118 $\pm$ 0.0351 & 0.258 $\pm$ 0.0316 & 0.276 $\pm$ 0.0743 & \underline{0.369 $\pm$ 0.2026} & n04044716          & 0.109 $\pm$ 0.0237       & 0.331 $\pm$ 0.0379       & 0.195 $\pm$ 0.0642       & \underline{0.363 $\pm$ 0.1572} \\
            n03590841          & 0.773 $\pm$ 0.0409 & 0.907 $\pm$ 0.0685 & 0.775 $\pm$ 0.0661 & \underline{0.597 $\pm$ 0.1035} & n04069434          & 0.091 $\pm$ 0.0299       & 0.217 $\pm$ 0.0381       & 0.139 $\pm$ 0.0985       & \underline{0.572 $\pm$ 0.1451} \\
            n03709823          & 0.279 $\pm$ 0.0295 & 0.304 $\pm$ 0.0668 & 0.296 $\pm$ 0.0736 & \underline{0.421 $\pm$ 0.2208} & n04086273          & 0.178 $\pm$ 0.0285       & 0.231 $\pm$ 0.0326       & 0.194 $\pm$ 0.0706       & \underline{0.505 $\pm$ 0.1369} \\
            n03773504          & 0.152 $\pm$ 0.0243 & 0.203 $\pm$ 0.0559 & 0.228 $\pm$ 0.0311 & \underline{0.616 $\pm$ 0.0932} & n04120489          & 0.038 $\pm$ 0.0102       & 0.159 $\pm$ 0.0307       & 0.059 $\pm$ 0.0471       & \underline{0.524 $\pm$ 0.2788} \\
            n03775071          & 0.207 $\pm$ 0.0147 & 0.291 $\pm$ 0.0217 & 0.213 $\pm$ 0.0544 & \underline{0.565 $\pm$ 0.1174} & n07753592          & 0.775 $\pm$ 0.0537       & 0.554 $\pm$ 0.0694       & 0.763 $\pm$ 0.0802       & \underline{0.894 $\pm$ 0.0487} \\
            n03792782          & 0.055 $\pm$ 0.0209 & 0.276 $\pm$ 0.0411 & 0.089 $\pm$ 0.0796 & \underline{0.524 $\pm$ 0.2397} & n07873807          & 0.371 $\pm$ 0.0259       & 0.347 $\pm$ 0.0401       & 0.316 $\pm$ 0.0989       & \underline{0.631 $\pm$ 0.0945} \\
            n03792972          & 0.085 $\pm$ 0.0277 & 0.259 $\pm$ 0.0379 & 0.114 $\pm$ 0.0645 & \underline{0.445 $\pm$ 0.1405} & n11939491          & 0.206 $\pm$ 0.0593       & 0.348 $\pm$ 0.0915       & 0.285 $\pm$ 0.0443       & \underline{0.606 $\pm$ 0.2026} \\ \cline{6-10}
            n03877472          & 0.104 $\pm$ 0.0186 & 0.123 $\pm$ 0.0137 & 0.165 $\pm$ 0.0718 & \underline{0.579 $\pm$ 0.1594} & {\it n13054560}    & {\it 0.000 $\pm$ 0.0000} & {\it 0.000 $\pm$ 0.0000} & {\it 0.000 $\pm$ 0.0000} & \underline{0.179 $\pm$ 0.0962} \\
            \bottomrule
        \end{tabular}
    }
\end{table*}

In addition to instance-level {\it retrieval}, we compare EEGVis-CMR with state-of-the-arts on semantic-level EEG-Visual {\it Classification}.
In this pipeline, a pre-trained visual classifier~(ResNet-50) is deployed to examine whether the model's visual output can be correctly recognized.
The output size of classifier is reduced to 40 as the classes of visual cues.
Although the classification metric was designed to evaluate image generators, we consider it reasonable to include EEGVis-CMR in the comparison from a task-oriented point of view; ultimately, an EEG-vision model should produce images that match the visual stimuli accurately.
We extend the evaluation to an open-set setting.
In detail, we select 38 subsets from the EEG dataset for model training, but excluding data from class {\tt n13054560}, which is then used as the open class in evaluation.
In this way, each approach only has access to EEG data and image data of 39 visual classes.
The evaluation is composed of a) closed-set classification on 39 known classes; and b) open-set detection for a novel class.

\subsubsection{Visual features facilitate Cross-modal learning.}
We report the EEG-visual {\it classification} results of 39 closed-set visual classes in Table~\ref{tab: classification}.
In {\it reconstruction}-based approaches, each EEG clip is encoded in regard to a semantic visual class, then conditioned to the decoder for image generation.
Brain2Image produces the lowest performance among four compared approaches, with slight edge between its results and random guesses for some classes.
One possible explanation is, the single-modal setup is used therein rather than cross-modal learning.
Thus, EEG encodings are solely derived from class supervision but not visual features of the stimuli.
NeuroVision focuses on improving the training of decoder with a progressively growth GAN, yielding more distinguished output than Brain2Image.
Still, visual features were neither included in EEG encoding, nor conditioned to the GAN training process.
This introduces artifacts into the synthesized images, and does not gain significant performance improvement as well.
Instead, DCLS-GAN concatenates EEG encoding with visual encoding as condition for the decoder.
An accuracy improvement of 11.8\% and 6.2\% over Brain2Image and NeuroVision can be found, implying that the decoding step could also benefit from visual modality.
Even so, both EEG and image encodings cannot provide ground-truth supervision like class labels to the decoder, causing incorrectly recognized visual outputs.
In contrast, EEGVis-CMR does not learn EEG encodings from class supervision, nor suffers from the quality of synthesized images but retrieves true visual cues instead.
It performs competitively in semantic-level EEG-visual {\it classification}, based on instance-level discriminative EEG encodings obtained from visual encodings.
It exhibits improved mean accuracy by 34.6\%, 22.7\% and 28.9\% over Brain2Image, DCLS-GAN, and NeuroVision, even without using class information.

\subsubsection{Self-supervision benefits open-set recognition.}

We discuss the merits of self-supervision upon performing open-set recognition.
Neither of three supervised {\it reconstruction}-based approaches can generate images outside the training classes, since a) supervised models assume test set contains the same classes as training set; and b) lack of a measure to detect novel class automatically.
In opposed, EEGVis-CMR recognizes similarity between EEG and image instances, and performs cross-modal retrieval.
Moreover, EEGVis-CMR applies contrastive learning to create a multi-modal representation space, satisfying alignment and uniformity~\cite{wang2020understanding}.
Then the cross-modal retrieval takes place upon instance-level similarity comparisons, rather than estimating the likelihood of belonging to a particular known semantic class, as in supervised counterparts.
EEGVis-CMR is shown to output images for {\tt n13054560} without touching class-level information.
This suggests that self-supervision may enhance implicit awareness of novel classes in this cross-modal learning task, even without auxiliary mechanisms for detecting novel classes.

\subsection{Discussion: Reconstruction vs Retrieval}
The {\it reconstruction}-based approaches rely on discriminative EEG encoder and powerful decoder to generate recognizable visual output.
The fact that encoders and decoders can be improved with either visual features or advanced training techniques, however, does not yet guarantee stable and high-quality image generation.
In addition, these methods are restricted to seen visual classes only due to supervised training pipeline, and thus fail at handling EEG signals and image data of novel classes.
In light of this, {\it retrieval}-based approach explores a different cross-modal learning paradigm, obtaining distinguishable visual output without training a generative decoder.
Moreover, self-supervised scheme creates a joint encoding space and uses visual features to train EEG encodings, providing instance-level supervision based on similarity calculation of multi-modal encodings.
Then cross-modal retrieval with a query can be made regardless of whether it belongs to a known class.
This avoids assigning novel unseen data to a known class,
eventually leading to improved model generalization.

\section{Conclusion}

We studied a self-supervised cross-modal learning paradigm to retrieve exact visual matching for an EEG clip evoked by visual stimulus, by maximizing the mutual information between correct EEG-image pairs.
We examined that self-supervised {\it retrieval} can overcome the limitations of visual classes and reproduce recognizable instances of visual stimuli accurately, while supervised {\it reconstruction} fail to do so.
Future works will extend this paradigm to other datasets and a wider range of subjects.

\clearpage
\bibliography{aaai23}

\end{document}